\documentclass[a4paper,fleqn]{cas-sc}

\usepackage[authoryear,longnamesfirst]{natbib}
\usepackage{booktabs}
\usepackage{amsmath}
\usepackage{amssymb}
\usepackage{graphicx}
\usepackage{hyperref}
\usepackage{tikz}
\usetikzlibrary{arrows.meta,positioning,fit,backgrounds,shapes.geometric}

\begin{document}
\let\WriteBookmarks\relax
\def\floatpagepagefraction{1}
\def\textpagefraction{.001}

\shorttitle{What survives honest evaluation?}
\shortauthors{Gençay}

\title[mode=title]{What survives honest evaluation? Leakage-safe, search-aware assessment of LLM-driven trading strategy discovery}

\author[1]{Eray Gen\c{c}ay}[orcid=0000-0002-1510-5628]
\ead{eray.gencay@gmail.com}

\affiliation[1]{
    organization={Independent Researcher},
    city={Stuttgart},
    country={Germany}
}

\begin{abstract}
Large language models (LLMs) are increasingly used to discover trading strategies, and much of the resulting literature shares a methodological weakness: many candidate strategies are generated, the best is reported, and neither look-ahead bias nor the intensity of the search behind the reported result is corrected for. We present a strategy-discovery system that makes both corrections structural rather than procedural. First, the agent can only act through registry-validated tools whose feature space excludes look-ahead by construction; we show that this guardrail is not redundant with statistical correction: a deliberately leaky oracle posting a Sharpe ratio of 35 \emph{survives} Deflated Sharpe and probability-of-backtest-overfitting testing completely. Second, the system records every strategy evaluation its search performs and deflates all reported performance by that trial count, tracing how the best in-sample Sharpe ratio climbs with each trial while the deflation threshold, driven by the agent's own search, climbs faster. Across a 453-stock point-in-time US equity universe and a 39-ETF multi-asset universe with realistic transaction, impact, and borrow costs, honest evaluation certifies passive benchmarks (out-of-sample confidence intervals excluding zero), rejects every LLM-discovered strategy (across two frontier models, search budgets up to one hundred candidates, and five repeated runs), catching selection luck, predicted rank degradation, and out-of-sample collapse through complementary instruments, and evaluates a human trader's production rule system under identical instruments. The framework formalizes why pre-registered hypotheses earn lower evidential bars than brute search, and quantifies the sample sizes that credible certification of moderate edges actually requires.
\end{abstract}

\begin{keywords}
large language models \sep algorithmic trading \sep backtest overfitting \sep deflated Sharpe ratio \sep data leakage \sep agent evaluation
\end{keywords}

\maketitle

\section{Introduction}
\label{sec:intro}

Large language models are now routinely embedded in systems that generate trading strategies: as alpha miners, as trading agents with memory and tool use, and as interactive co-pilots for quantitative research \citep{YangFinGPT2023,YuFinMem2024,ZhangFinAgent2024,WangAlphaGPT2023,WangQuantAgent2024,YuAlphaGen2023}. This literature has grown quickly, and it has inherited a methodological weakness that the empirical-finance community diagnosed years before LLMs arrived: when a search process generates many candidate strategies and the best backtest is reported, the reported performance is inflated by selection, and small violations of point-in-time discipline inflate it further \citep{Bailey2014,BaileyLdP2014dsr,Harvey2016,HarveyLiu2015backtesting,Arnott2019,LopezdePrado2018}. The instruments to correct both pathologies exist (the Deflated Sharpe Ratio, the probability of backtest overfitting, strict point-in-time data protocols), but they are rarely applied in the LLM strategy-discovery literature, a gap that independent audits have since documented \citep{Yao2026beyond,Li2025finsaber}. The result is a growing collection of reported Sharpe ratios that a skeptical practitioner cannot readily act on.

An autonomous agent makes this problem worse in a specific, quantifiable way: \emph{autonomy inflates the trial count}. An LLM agent that proposes, evaluates, and refines strategies in a loop performs dozens of implicit backtests before its operator sees a single number. If the reported result is not deflated by that search intensity, the agent's very productivity becomes a bias amplifier. Conversely, prompting an LLM to ``avoid look-ahead bias'' is a weak guardrail: a text instruction does not prevent a generated strategy from consuming future information if the execution environment exposes it.

This paper's position is that both corrections must be \emph{structural}, properties of the system that hold by construction, rather than procedural conventions that authors promise to have followed. We present a strategy-discovery system built around two such structures and an experimental protocol that tests what remains of LLM-discovered alpha once they are in place. We report the results symmetrically: what the evaluation rejects \emph{and} what it certifies. On the most heavily arbitraged corner of world markets, no active strategy in our suite survives, while the same instruments certify passive risk premia, expose a planted look-ahead oracle as uncatchable by deflation alone, predict an overfit discovery before it fails out-of-sample, and give a human trader's production system the same hearing as the LLM's proposals. Although the instruments come from empirical finance, the underlying selection problem is not specific to it: any agentic system that searches a solution space and self-reports its best result faces the same inflation. The design pattern demonstrated here, a validated action surface plus a complete self-recorded search ledger, is offered as an evaluation architecture for autonomous discovery; trading is the case study in which the corrective instruments already exist.

Our contributions are:

\begin{enumerate}
\item \textbf{A leakage-safe discovery architecture.} The agent composes strategies exclusively through registry-validated tools: every feature, signal-graph node, and portfolio archetype it can reference is checked in-loop, parameters are validated against typed schemas, and the feature registry excludes look-ahead constructions from the agent-selectable set. Look-ahead is not discouraged; it is \emph{inexpressible}. A deterministic audit layer independently enforces publication delays for external event data. The architecture includes a stateful rule-based archetype that can express discrete human trading systems (entry/exit rules with position state), which we use to port a production strategy rule-for-rule.
\item \textbf{Search-aware honest evaluation.} The system records every distinct strategy evaluation its search performs in a trial ledger and computes, for every reported arm: the Deflated Sharpe Ratio against the search's own trial count and Sharpe dispersion \citep{BaileyLdP2014dsr}, the probability of backtest overfitting via combinatorially symmetric cross-validation \citep{Bailey2014}, stationary-bootstrap confidence intervals \citep{PolitisRomano1994}, and paired tests against buy-and-hold. An \emph{evaporation curve} traces the best in-sample Sharpe ratio against its rising deflation threshold as the agent's trial count grows.
\item \textbf{A six-experiment protocol on real data.} E1--E6b each answer a named methodological objection, on a 453-stock point-in-time US equity universe and a 39-ETF multi-asset universe, with realistic costs (commissions, spreads, square-root market impact, short borrow) and held-out evaluation windows of four and nine years.
\item \textbf{Findings with practical import.} (i) Deflation does not catch leakage: a planted oracle with a Sharpe ratio of 35 passes every statistical test, so the two guardrails are complementary, not substitutes. (ii) Honest evaluation certifies passive premia while rejecting every discovered strategy, across universes, cost levels, two frontier LLMs, and search budgets up to one hundred candidates; three distinct failure modes (selection luck; predicted rank degradation; out-of-sample collapse) are each caught by the instrument built for them. (iii) The framework formalizes why a pre-registered hypothesis faces almost no deflation while a broad search faces a high bar, and the statistical-power arithmetic quantifies why certifying moderate low-frequency edges requires decades of out-of-sample data or substantial breadth.
\end{enumerate}

The remainder of the paper is organized as follows. Section~\ref{sec:related} positions the work between the LLM-agent finance literature and the backtest-overfitting literature. Section~\ref{sec:system} describes the discovery architecture. Section~\ref{sec:honest} describes the evaluation methodology. Section~\ref{sec:design} specifies the experimental protocol, and Section~\ref{sec:results} reports the results. Section~\ref{sec:discussion} discusses implications, Section~\ref{sec:limitations} states limitations, and Section~\ref{sec:conclusion} concludes.

\section{Related work}
\label{sec:related}

\paragraph{LLMs and agents for trading.}
A rapidly growing body of work applies LLMs to strategy generation and trading. FinGPT provides open financial LLM infrastructure \citep{YangFinGPT2023}; FinMem and FinAgent build trading agents with layered memory, multimodal inputs, and tool augmentation \citep{YuFinMem2024,ZhangFinAgent2024}; Alpha-GPT and QuantAgent frame alpha mining as human--AI interaction or self-improving search \citep{WangAlphaGPT2023,WangQuantAgent2024}; AlphaGen searches formulaic alpha collections with reinforcement learning \citep{YuAlphaGen2023}. These systems demonstrate impressive generation capability, and they typically evaluate discovered strategies by reporting backtested performance, in some cases with out-of-sample splits, but, to our knowledge, never with multiple-testing correction indexed to the search's own trial count, and never with a demonstration that their execution environments make look-ahead impossible rather than merely discouraged. Earlier work on machine learning for asset pricing established both the promise and the fragility of data-driven signals \citep{Gu2020,Heaton2017,Fischer2018,Kelly2019,Dixon2020}, and text-based signals have a long pre-LLM history \citep{Tetlock2007,Tetlock2008,Loughran2011,Engelberg2011,Bollen2011,Ranco2015,Nassirtoussi2014,Luss2015,Yang2020,Wu2023}.

\paragraph{Backtest overfitting and multiple testing.}
A parallel literature, largely disjoint from the above, quantifies how search inflates backtests. \citet{Bailey2014} define the probability of backtest overfitting (PBO) and estimate it by combinatorially symmetric cross-validation (CSCV); \citet{BaileyLdP2014dsr} derive the Deflated Sharpe Ratio (DSR), which tests an observed Sharpe ratio against a threshold that rises with the number of trials and their dispersion; \citet{Harvey2016} and \citet{HarveyLiu2015backtesting} argue that most claimed factors fail appropriately deflated hurdles; \citet{Arnott2019} codify a backtesting protocol; \citet{LopezdePrado2018} treats leakage, embargoing, and deflation as first-class concerns. This literature supplies exactly the instruments the LLM literature lacks, but it predates agents, and it has no notion of a system that \emph{records its own search} so that deflation can be indexed to it.

\paragraph{Tool-augmented agents.}
Tool use is usually motivated by capability: letting models call calculators, retrievers, or APIs \citep{Schick2023,Yao2023,Wei2022}. Multi-agent decompositions assign specialized roles \citep{Park2023,Li2023CAMEL,Hong2023MetaGPT,Wooldridge2009}. We use the same machinery for a different end: a \emph{validated tool surface as a safety boundary}. The agent's action space is the registry of typed, audited operations; anything outside it, including any look-ahead feature, is not a possible action. Interpretability requirements in high-stakes settings \citep{Rudin2019,Lo2010} favor the same design, since every discovered strategy is a declarative, human-readable plan rather than generated code.

\paragraph{Independent audits and re-evaluations.}
Two contemporaneous works document the same gap from outside our harness. \citet{Yao2026beyond} audit execution assumptions across thirty trade-relevant LLM studies and conclude that system architecture is reported more consistently than the evaluation assumptions a reader needs in order to judge whether a result is interpretable; they ship a reporting checklist, but they do not re-evaluate the systems they audit. \citet{Li2025finsaber} re-evaluate published LLM timing strategies over two decades and more than a hundred symbols and find that the reported advantages deteriorate substantially under the broader cross-section and longer horizon, naming survivorship, look-ahead, and data-snooping bias explicitly. Their remedy is to broaden the evaluation until inflated advantages wash out; ours is to make the search intensity itself the deflator, so a result is judged against the number of trials that produced it. Neither work makes leakage-safety a property of the action space, and neither deflates a reported number by a recorded trial count, but both are independent evidence that the pattern this paper corrects for is real.

\paragraph{Gap.}
We are not aware of prior work that connects these literatures: an LLM discovery system whose leakage-safety is structural, whose search intensity is recorded by construction, and whose every reported number carries the corresponding deflated verdict. The audits above measure the gap from the outside; this paper addresses it inside a discovery system.

\section{A discovery system that is honest by construction}
\label{sec:system}

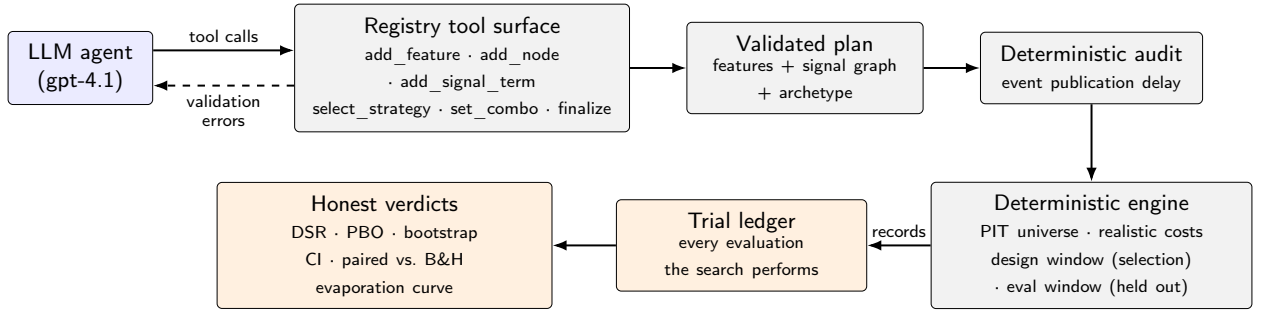
\begin{figure*}[pos=t]
\centering
\resizebox{\linewidth}{!}{%
\begin{tikzpicture}[
  font=\small,
  box/.style={draw, rounded corners=2pt, align=center, inner sep=5pt, minimum height=9mm},
  agent/.style={box, fill=blue!8},
  det/.style={box, fill=gray!10},
  eval/.style={box, fill=orange!12},
  arr/.style={-{Latex[length=2mm]}, thick},
  lbl/.style={font=\scriptsize, midway}
]
\node[agent, text width=17mm] (llm) {LLM agent\\(gpt-4.1)};
\node[det, right=20mm of llm, text width=44mm] (tools)
  {Registry tool surface\\
   {\scriptsize add\_feature\ \textperiodcentered\ add\_node\ \textperiodcentered\ add\_signal\_term}\\
   {\scriptsize select\_strategy\ \textperiodcentered\ set\_combo\ \textperiodcentered\ finalize}};
\node[det, right=8mm of tools, text width=30mm] (plan)
  {Validated plan\\{\scriptsize features + signal graph\\+ archetype}};
\node[det, right=8mm of plan, text width=28mm] (audit)
  {Deterministic audit\\{\scriptsize event publication delay}};
\node[det, below=11mm of audit, text width=42mm] (engine)
  {Deterministic engine\\
   {\scriptsize PIT universe\ \textperiodcentered\ realistic costs}\\
   {\scriptsize design window (selection)\ \textperiodcentered\ eval window (held out)}};
\node[eval, left=9mm of engine, text width=32mm] (ledger)
  {Trial ledger\\{\scriptsize every evaluation\\the search performs}};
\node[eval, left=9mm of ledger, text width=44mm] (verdict)
  {Honest verdicts\\
   {\scriptsize DSR\ \textperiodcentered\ PBO\ \textperiodcentered\ bootstrap CI\ \textperiodcentered\ paired vs.\ B\&H}\\
   {\scriptsize evaporation curve}};
\draw[arr] ([yshift=2.5mm]llm.east) -- node[lbl, above]{tool calls} ([yshift=2.5mm]tools.west);
\draw[arr, dashed] ([yshift=-2.5mm]tools.west) -- node[lbl, below, align=center]{validation\\errors} ([yshift=-2.5mm]llm.east);
\draw[arr] (tools) -- (plan);
\draw[arr] (plan) -- (audit);
\draw[arr] (audit) -- (engine);
\draw[arr] (engine) -- node[lbl, above]{records} (ledger);
\draw[arr] (ledger) -- (verdict);
\end{tikzpicture}%
}
\caption{System overview. The agent acts only through registry-validated tools; invalid calls return structured errors for in-loop self-correction, and look-ahead features are not in the agent-selectable registry. Validated plans pass a deterministic audit and are executed by a deterministic engine over a point-in-time universe with realistic costs, backtesting separately on the design window (where all selection happens) and the held-out evaluation window. Every evaluation the search performs is recorded in the trial ledger, which drives the deflated verdicts attached to every reported number.}
\label{fig:architecture}
\end{figure*}

\subsection{Strategies as declarative plans}
A strategy in our system is a \emph{plan}: a list of typed feature specifications (e.g., \texttt{momentum(window=252)}), an optional signal graph of named nodes (transforms over features or earlier nodes: arithmetic, comparisons, rolling statistics, lags with non-negative shift only), a weighted signal combination, and a portfolio \emph{archetype} that turns the signal into target weights. Five archetypes are registered: a linear tilt; a cross-sectional rank long/short (top-$k$/bottom-$k$, equal-weighted, with a re-ranking cadence); a volatility-targeting overlay that scales a base archetype's weights to a trailing realized-volatility target; a deterministic random-weights null; and a stateful rule engine (Section~\ref{sec:rulearch}). Plans are declarative and human-readable; every experiment in this paper, including the port of a human trader's production system, is a JSON configuration interpretable at the feature level. This satisfies the interpretability desideratum directly \citep{Rudin2019}: there is no generated code to audit, only named features and named rules.

\subsection{The tool surface as a safety boundary}
\label{sec:tools}
The agent never writes code and never emits free-form strategy descriptions that are later parsed on trust. It composes a plan by calling tools generated from the same registries that define the system (Fig.~\ref{fig:architecture}): \texttt{add\_feature}, \texttt{add\_node}, \texttt{add\_signal\_term}, \texttt{select\_strategy}, \texttt{set\_combo}, \texttt{finalize}. Each call is validated immediately against parameter schemas, reference resolution, archetype names, and portfolio-knob allow-lists; a signal term may only consume features or nodes that exist. An invalid call returns a structured error that the model corrects in-loop. Three properties follow.

First, \emph{look-ahead is inexpressible}. The feature registry contains a deliberately leaky family (used in experiment E1), but it is excluded from the agent-selectable set; the tool schema's enumeration, the catalog shown to the model, and the server-side validation all derive from the same registry flag. Second, \emph{parameter hallucination is eliminated rather than punished}. In a matched 60-iteration run of a JSON-emitting critic without in-loop validation, 60 of 60 proposals were rejected for invalid feature names or parameters; through the tool loop, 99 of 100 (gpt-4.1) and 80 of 100 (claude-sonnet-5) proposals in the hundred-candidate searches of Section~\ref{sec:e3} were valid, because each feature's exact parameters are advertised in the tool catalog and validated per call; every invalid construction (duplicate feature roles, empty plans) was rejected by validation rather than evaluated. Third, \emph{the search becomes observable}: every candidate the agent composes flows through the same evaluation entry point, which is what makes the trial ledger of Section~\ref{sec:honest} complete by construction rather than by author diligence.

A deterministic audit layer independently checks event-timing declarations (external signals must respect a publication delay), and rejects plans rather than repairing them. Execution is handled by a deterministic engine: signals computed at day $t$ use bars up to and including $t$ and are held from $t{+}1$ (the engine applies the one-day execution lag globally); costs include commissions, spreads, square-root market impact against trailing dollar volume, and short-borrow financing; the tradeable universe is re-selected point-in-time from trailing liquidity, so universe membership embeds no future information.

\paragraph{Relation to structured generation.} Registry-validated tool composition is complementary to, not in competition with, grammar- and schema-constrained decoding \citep{WillardLouf2023,Geng2023grammar} or typed function calling. Those techniques enforce \emph{syntactic} validity, meaning the output parses and the types check, and the tool surface here already inherits that layer, since each tool is a typed call whose enumerations derive from the registries. What token-level constraints cannot enforce is validity that depends on system state and accumulated plan context: whether a feature is registered \emph{and} flagged leakage-safe, whether a reference resolves to a node the agent has actually added, whether a parameter satisfies its registered range. Those checks live server-side, in the same validation path that human-authored configurations traverse, and their structured error returns are what let the model repair invalid calls in-loop. The load-bearing property is not the shape of the model's output but that validity is defined once, by the system's registries, and enforced at the single entry point that also makes the trial ledger complete.

\subsection{Autonomous discovery}
\label{sec:autonomous}
Discovery runs as a bounded, design-only refinement loop. In each iteration the LLM composes a complete fresh strategy through the tool loop under a rotating economic hypothesis hint (momentum, reversal, low volatility, volatility breakout, volume, oscillator mean-reversion, trend filters, and combinations); each candidate is audited, evaluated on the design window only, and recorded. The evaluation window is never visible to the search. The manifest records the model identity (\texttt{gpt-4.1} for the primary experiments, with a cross-model replication on \texttt{claude-sonnet-5}; see Section~\ref{sec:e3}), the full per-iteration trace, and the resulting trial ledger.

\subsection{A rule archetype for human systems}
\label{sec:rulearch}
Human traders rarely hold continuous signal-proportional positions; they enter when a rule fires, hold with position state, and exit on a different rule. A stateless weight map cannot express this hysteresis. The \texttt{rule\_long\_short} archetype reads four named condition nodes from the plan's signal graph (\texttt{long\_entry}, \texttt{long\_exit}, \texttt{short\_entry}, \texttt{short\_exit}) and walks dates carrying per-symbol position state: a flat book enters when an entry condition fires (blocked if the same side's exit fires on the same bar, matching common trading-platform semantics), and holds until the exit condition fires; a decision cadence parameter lets a weekly-chart system run on daily bars at its native rhythm. Unknown (warm-up) conditions never trigger actions. This archetype is exercised in experiment E6b, where a production PineScript system, a stateful trend-following rule engine with candle-pattern filters, is expressed as an 84-node signal graph and cross-validated, bar for bar, against a reference implementation of the original script.

\section{Search-aware honest evaluation}
\label{sec:honest}

Every reported number in this paper carries a deflated verdict. This section defines the instruments; all are computed per run and persisted in the run's manifest.

\subsection{Deflated Sharpe Ratio indexed to the recorded search}
Let $\widehat{SR}$ be the per-observation Sharpe ratio of a strategy's \emph{design-window} returns over $T$ observations with skewness $\gamma_3$ and kurtosis $\gamma_4$. The Probabilistic Sharpe Ratio against a benchmark $SR^{*}$ is
\begin{equation}
\mathrm{PSR}(SR^{*}) \;=\; \Phi\!\left(\frac{(\widehat{SR}-SR^{*})\sqrt{T-1}}{\sqrt{1-\gamma_3 \widehat{SR} + \tfrac{\gamma_4-1}{4}\widehat{SR}^{2}}}\right),
\end{equation}
and the Deflated Sharpe Ratio is $\mathrm{DSR} = \mathrm{PSR}(SR^{*}_{0})$ with the selection-aware threshold
\begin{equation}
SR^{*}_{0} \;=\; \sqrt{V}\left[(1-\gamma)\,\Phi^{-1}\!\big(1-\tfrac{1}{N}\big) + \gamma\,\Phi^{-1}\!\big(1-\tfrac{1}{Ne}\big)\right],
\label{eq:dsr-threshold}
\end{equation}
where $N$ is the number of trials, $V$ the variance of the trials' Sharpe ratios, and $\gamma$ the Euler--Mascheroni constant \citep{BaileyLdP2014dsr}. The crucial design decision is what $N$ and $V$ are: in this system they come from a \emph{trial ledger} that records every distinct strategy evaluation the search performed, meaning every refinement iteration that produced a design-window backtest and every experimental arm, rather than from an author's estimate. Because all agent-composed candidates flow through one evaluation entry point (Section~\ref{sec:tools}), the ledger is complete by construction. Deliberately planted demonstration arms (the E1 oracle) are excluded so that their pathological Sharpe ratios cannot distort the deflation of honest arms.

\subsection{Probability of backtest overfitting}
PBO is estimated by CSCV \citep{Bailey2014}: the design window is partitioned into $S{=}16$ blocks; for each of the $\binom{S}{S/2}$ in-sample/out-of-sample block combinations, the in-sample-best candidate's out-of-sample rank $\omega$ yields a logit $\lambda = \ln\frac{\omega}{1-\omega}$, and $\mathrm{PBO} = P(\lambda \le 0)$, the probability that the search's in-sample winner underperforms the median candidate out of sample. The candidate matrix contains the design-window return series of every ledger entry, so the column count equals the trial count by an enforced invariant.

\subsection{Out-of-sample inference}
On the held-out evaluation window we report stationary-bootstrap confidence intervals \citep{PolitisRomano1994} for annualized Sharpe ratio and total return, and a paired block-bootstrap test of daily return differences against buy-and-hold. Design-window instruments (DSR, PBO) measure selection; evaluation-window instruments measure what survived it. A strategy is \emph{certified} in our terminology when its evaluation-window Sharpe confidence interval excludes zero.

\subsection{The evaporation curve}
\label{sec:evaporation}
For $k = 1,\dots,N$ over the ledger in evaluation order, take the best-of-first-$k$ candidates by design Sharpe ratio and deflate it by $k$ trials and the dispersion of those $k$ Sharpe ratios (Eq.~\ref{eq:dsr-threshold}). The resulting curve shows the raw best Sharpe ratio (a running maximum, non-decreasing) against its deflation threshold (increasing in $k$ and dispersion). Where the threshold overtakes the best, the search's own intensity has outrun its discoveries: the marginal trial costs more evidence than it finds. This makes the multiple-testing accounting \emph{visible} in a single figure (Fig.~\ref{fig:evaporation}).

\subsection{Pre-registration semantics}
Equation~\ref{eq:dsr-threshold} has an important corollary: at $N{=}1$ the threshold collapses to the benchmark and DSR equals PSR, so a single pre-registered hypothesis faces almost no deflation. The framework therefore \emph{rewards} hypothesis-driven design over brute search, formalizing why a single pre-specified economic thesis tested against decades of evidence earns a lower evidential bar than a machine that tried a thousand variants, unless the machine pays for the thousand. A human hypothesis enters the protocol as a pre-registered arm (E6/E6b) and the ledger grows honestly with each specified variant.

\section{Experimental design}
\label{sec:design}

The suite is organized so that each experiment answers a specific methodological objection (Table~\ref{tab:map}).

\begin{table}[pos=t]
\centering\small
\caption{The objection-to-experiment map. Each experiment answers a methodological objection a skeptical reviewer or practitioner would raise.}
\label{tab:map}
\begin{tabular}{@{}lp{0.62\columnwidth}@{}}
\toprule
Exp. & Objection answered \\
\midrule
E1 & ``Leakage-safe is asserted, not shown. And doesn't deflation catch leakage anyway?'' \\
E2 & ``Does anything simple survive honest evaluation on this universe? What does an honest factor search show?'' \\
E3 & ``What happens when the LLM itself discovers strategies, and its own search is the deflator?'' \\
E4 & ``Is the conclusion an artifact of cost assumptions, universe breadth, or one LLM sample?'' \\
E5 & ``Is this just an expensive proof of the random walk? Can the referee ever say yes?'' \\
E6/E6b & ``How does a human trader's pre-registered production system fare under identical instruments?'' \\
\bottomrule
\end{tabular}
\end{table}

\paragraph{Data and universes.}
E1--E4 run on daily OHLCV bars for 453 liquid US large-capitalization stocks plus the SPY benchmark (Tiingo, 2015--2026). The tradeable set is re-selected point-in-time as the top 200 names by trailing 63-day dollar volume, re-evaluated every 21 trading days, which removes universe-selection look-ahead. The constituent list is a fixed current list; the resulting survivorship bias is disclosed and bounded in Section~\ref{sec:limitations} (it flatters active strategies, making our null conservative), and the panel does include eight names that delist mid-sample. E5--E6b run on 39 multi-asset ETFs (country and sector equity, Treasuries and credit, gold and commodities, currencies, bitcoin; 2005--2026), where documented cross-asset premia reside \citep{Moskowitz2012tsm,Asness2013,Jegadeesh1993}.

\paragraph{Windows and power.}
E1--E4 use design 2017--2021 and held-out evaluation 2022--2025. E5--E6b use design 2007--2016 and evaluation 2017--2025: the nine-year window is a deliberate power choice, since a $t$-statistic on a Sharpe ratio grows as $SR\sqrt{\text{years}}$, so nine years suffices to certify $SR \approx 0.7$ at conventional levels while four years cannot certify moderate edges at all, a fact the results repeatedly illustrate.

\paragraph{Costs.}
All experiments charge 1\,bp commission, 2\,bp spread, square-root market impact against trailing 21-day dollar volume, and 50\,bp annualized borrow on short notional. E4 stress-tests these at $\times 0.5$ and $\times 2$.

\paragraph{Model and reproducibility.}
Agent runs use OpenAI \texttt{gpt-4.1}, with a cross-model replication on Anthropic \texttt{claude-sonnet-5}, through a provider-agnostic interface; the manifest records provider, model, code revision, package versions, data content hashes, the full refinement trace, and the statistics block. Every experiment is a JSON configuration; the deterministic experiments (E1, E2, E4 cost/universe variants, E5 reference arms, E6, E6b) involve no LLM calls and reproduce exactly.

\section{Results}
\label{sec:results}

\subsection{E1: deflation does not catch leakage}
\label{sec:e1}

A deliberately leaky feature (tomorrow's return, available today) is registered in the system but excluded from the agent-selectable set; E1 plants it in a literal configuration arm, something the agent path cannot do, next to a leakage-safe momentum arm, a random-weights null, and passive benchmarks. The oracle is a proof by counterexample: an intentionally extreme contamination that bounds what any statistics-only defense can catch, not a model of the subtler alignment errors that produce leakage in practice.

\begin{table}[pos=t]
\centering\small
\caption{E1 (453-stock universe, PIT top-200; design 2017--2021, eval 2022--2025). The look-ahead oracle survives statistical deflation completely; only the structural guardrail removes it. The oracle is excluded from the trial ledger so its Sharpe cannot distort other arms' deflation.}
\label{tab:e1}
\begin{tabular}{@{}lrrr@{}}
\toprule
Arm & Design SR & Eval SR & DSR \\
\midrule
Look-ahead oracle (unsafe) & 34.7 & 51.5 & \textbf{1.00} \\
Buy-and-hold (equal-weight) & 1.15 & 0.60 & 0.99 \\
SPY buy-and-hold & 0.98 & 0.67 & 0.97 \\
Random floor & 0.06 & $-$0.09 & 0.46 \\
Momentum L/S (safe) & $-$0.07 & 0.01 & 0.35 \\
\bottomrule
\end{tabular}
\end{table}

Table~\ref{tab:e1} shows the central negative result for statistics-only defenses: the oracle posts a design Sharpe ratio of 34.7, an \emph{evaluation} Sharpe of 51.5, and a DSR of 1.00. Deflation corrects for \emph{selection among honestly computed backtests}; it has no mechanism against a strategy whose information set is contaminated, and the contamination survives out of sample by construction. Leakage-safety and search-deflation are therefore complementary guardrails rather than substitutes, and a framework relying on deflation alone leaves this failure mode uncovered. The magnitude of the oracle's Sharpe also calibrates what leakage \emph{buys}: a reviewer seeing a two-digit Sharpe should suspect the information set, not the alpha.

\subsection{E2: the baseline landscape}
\label{sec:e2}

E2 runs classic price factors (126-day momentum, low-volatility, short-term reversal, all as market-neutral rank long/short books) against the random null and passive benchmarks, with a small hyperparameter grid refined on the design window (trial ledger $N{=}9$).

\begin{table}[pos=t]
\centering\small
\caption{E2 baseline landscape (same universe/windows as E1). No classic factor survives deflation; the search itself is flagged as overfit (PBO $=0.83$).}
\label{tab:e2}
\begin{tabular}{@{}lrrr@{}}
\toprule
Arm & Design SR & Eval SR & DSR \\
\midrule
Buy-and-hold (equal-weight) & 1.15 & 0.60 & \textbf{0.97} \\
SPY buy-and-hold & 0.98 & 0.67 & 0.94 \\
Random floor & 0.06 & $-$0.09 & 0.32 \\
Short-term reversal L/S & $-$0.04 & $-$0.20 & 0.25 \\
Momentum L/S & $-$0.07 & 0.01 & 0.22 \\
Low-volatility L/S & $-$0.58 & $-$0.78 & 0.03 \\
\bottomrule
\end{tabular}
\end{table}

\begin{figure}[pos=t]
\centering
\includegraphics[width=\columnwidth]{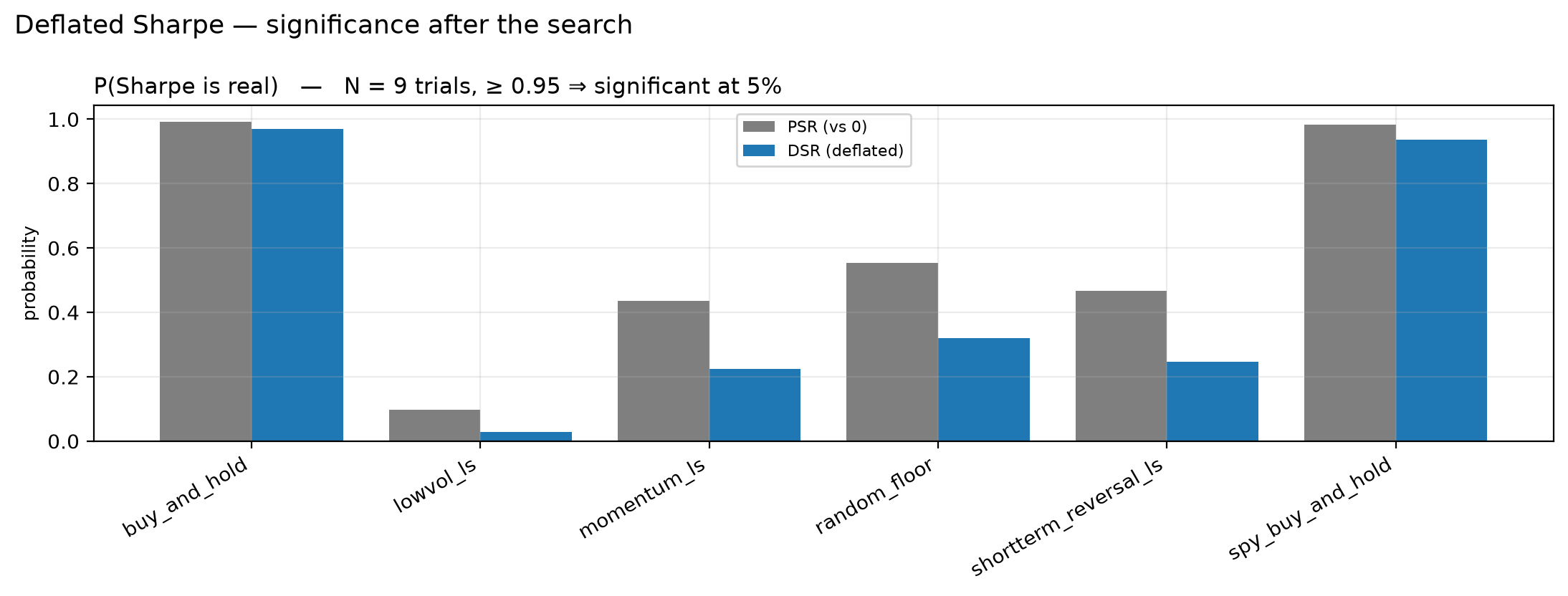}
\caption{E2: Probabilistic (grey) vs.\ Deflated (blue) Sharpe per arm. The gap between the bars is the price of the search; only equal-weight buy-and-hold clears the 0.95 significance level (SPY sits just below at 0.94), and every active arm is far away.}
\label{fig:dsr-e2}
\end{figure}

\begin{figure}[pos=t]
\centering
\includegraphics[width=\columnwidth]{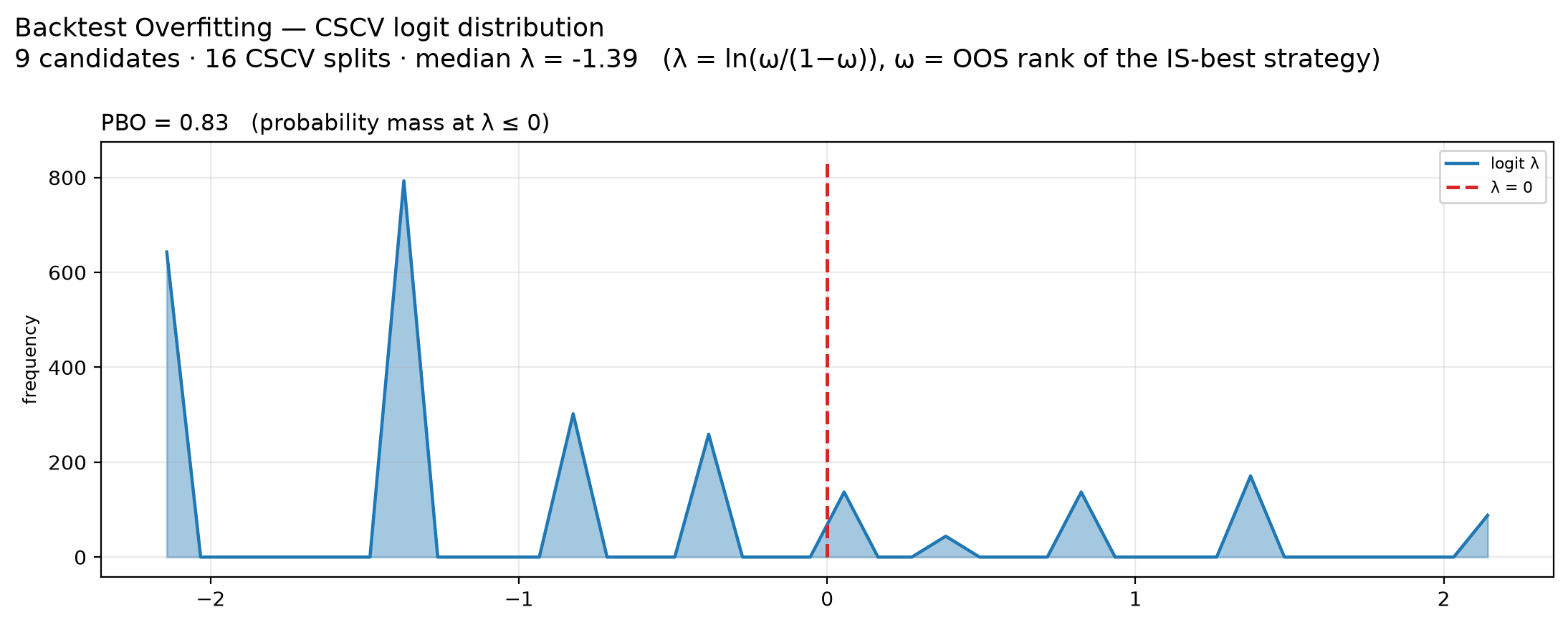}
\caption{E2: CSCV logit distribution. The mass at $\lambda \le 0$ is the probability that the in-sample-best candidate underperforms the median out of sample (PBO $=0.83$).}
\label{fig:pbo-e2}
\end{figure}

No factor comes near significance (Table~\ref{tab:e2}, Fig.~\ref{fig:dsr-e2}): every active arm sits at DSR $\le 0.32$, the CSCV estimate flags the hyperparameter search itself as likely overfit (PBO $= 0.83$; Fig.~\ref{fig:pbo-e2}), and only the passive arms approach the 0.95 level, where equal-weight buy-and-hold clears it (0.97) and SPY sits just below (0.94). On one of the most heavily arbitraged corners of world markets, with realistic costs and market-neutral construction that deliberately forfeits the equity risk premium, this is the baseline against which the discovery claims that follow should be read.

\subsection{E3: autonomous discovery, deflated by its own search}
\label{sec:e3}

E3 is the headline protocol. The agent composes one hundred complete strategies through the tool loop under a rotating set of twenty economic hypothesis hints (Section~\ref{sec:autonomous}); 99 of 100 candidates are valid, interpretable plans (versus 0 of 60 for a JSON-emitting critic without in-loop validation), spanning 22 distinct feature combinations (unique sets of feature names among the evaluated candidates; parameter values and signal wiring may differ within a set). The ledger reaches $N{=}102$.

\begin{table}[pos=t]
\centering\small
\caption{E3 autonomous discovery (gpt-4.1, $N{=}102$ trials, PBO $=0.01$). The agent's best find beats buy-and-hold in sample, then fails certification against its own trial count and collapses out of sample.}
\label{tab:e3}
\begin{tabular}{@{}lrrrr@{}}
\toprule
Arm & Design SR & Eval SR & Eval ret. & DSR \\
\midrule
Agent (RSI $\times$ volume) & \textbf{1.69} & 0.18 & $+4.7\%$ & \textbf{0.86} \\
Buy-and-hold & 1.15 & 0.60 & $+40.8\%$ & 0.45 \\
SPY buy-and-hold & 0.98 & 0.67 & $+51.5\%$ & 0.31 \\
Momentum reference & $-$0.07 & 0.01 & $-1.6\%$ & 0.00 \\
Random floor & 0.06 & $-$0.09 & $-1.0\%$ & 0.01 \\
\bottomrule
\end{tabular}
\end{table}

\begin{figure}[pos=t]
\centering
\includegraphics[width=\columnwidth]{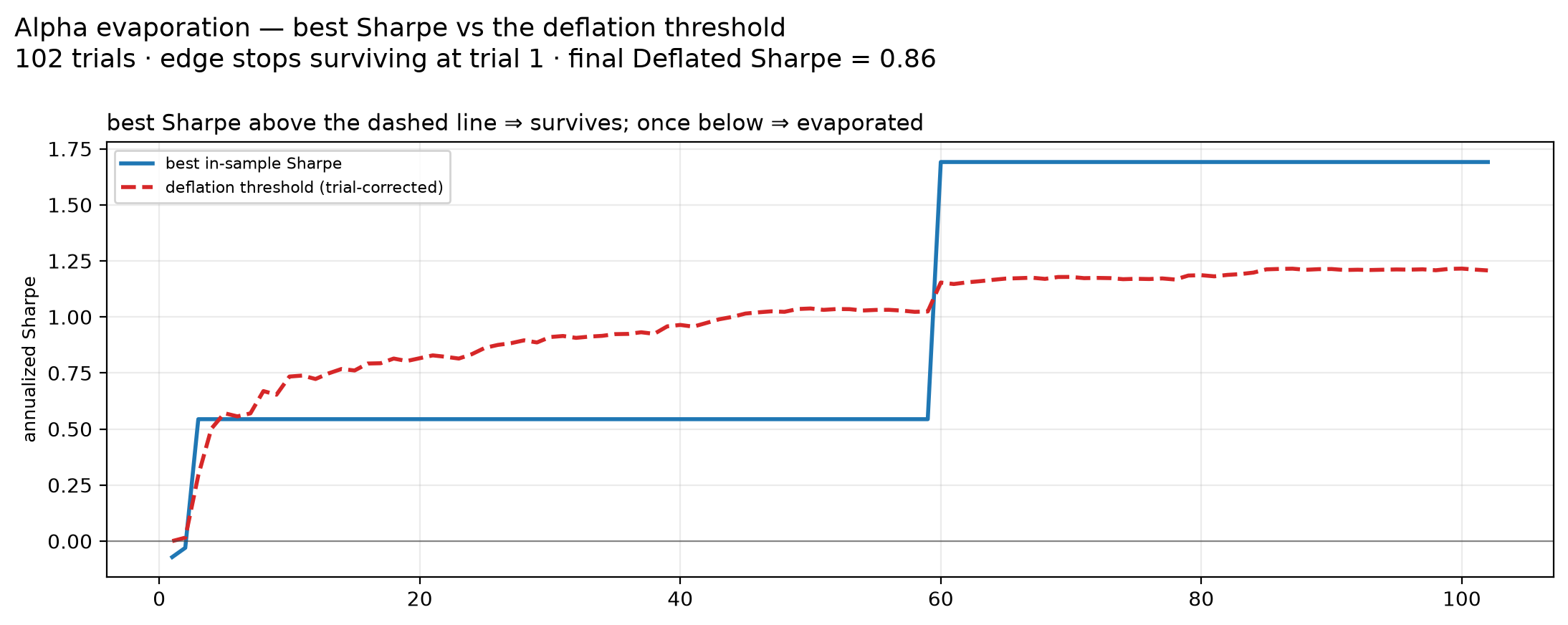}
\caption{E3: the evaporation curve over 102 self-recorded trials. The agent's best in-sample Sharpe (solid) is a running maximum, climbing from $-0.07$ to $1.69$; the deflation threshold implied by the agent's own trial count and Sharpe dispersion (dashed) chases it to $1.21$. The search partially outruns its bar (DSR ends at $0.86$) yet remains below the $0.95$ certification level, and the held-out window (Table~\ref{tab:e3}) vindicates the caution.}
\label{fig:evaporation}
\end{figure}

The result is deliberately uncomfortable (Table~\ref{tab:e3}). The agent's best discovery, an interpretable contrarian book over RSI and a volume z-score, re-ranked monthly, posts a design Sharpe of \textbf{1.69}, better in sample than buy-and-hold. It is the kind of number a paper could lead with. Honest evaluation withholds certification: after the agent's own 102 recorded trials, the deflation threshold reaches 1.21 and the DSR stops at 0.86, short of the 0.95 level (Fig.~\ref{fig:evaporation}), and the held-out window vindicates the caution, with the discovery collapsing to an evaluation Sharpe of 0.18 ($+4.7\%$ over four years, CI $[-0.80, +1.14]$) while buy-and-hold returned $+40.8\%$. Notably, PBO is \emph{low} here (0.01): the winner keeps its \emph{rank} among the mostly losing candidates out of sample while losing its absolute edge, so the failure is caught by the held-out interval, not the rank statistic (Section~\ref{sec:discussion}). Two presentation notes: the benchmark arms were not searched, so their deflation by this search's wide dispersion is conservative (as pre-registered references their $N{=}1$ verdicts are their PSRs, 0.99 and 0.98; Section~\ref{sec:honest}), and a grid-refined single-proposal variant of the protocol ($N{=}21$, PBO $=0.11$) reaches the same verdict from the other side: its discovery lost in both windows (design $-0.80$, evaluation $-0.75$, DSR $=0.001$).

\paragraph{Cross-model replication.} The identical protocol on a second vendor's frontier model, Anthropic's \texttt{claude-sonnet-5}, reaches the same verdict through different search behavior: 80 of 100 proposals valid (all twenty rejections are invalid constructions, sixteen duplicate feature roles and four empty plans, caught by validation and never evaluated), 39 distinct feature combinations explored (a more diverse search than gpt-4.1's 22), best find a price-to-moving-average spread with design Sharpe 0.44 that degrades out of sample to $-0.33$ ($-29\%$), DSR $=0.18$, PBO $=0.51$. Neither vendor's hundred-candidate search produces a certifiable strategy; each manifest records the provider, model, and complete search trace.

\subsection{E4: robustness of the null}
\label{sec:e4}

\begin{table}[pos=t]
\centering\small
\caption{E4 robustness. Top: deterministic variants of E2, reporting each variant's strongest factor. Bottom: five independent repetitions of the E3 discovery protocol.}
\label{tab:e4}
\begin{tabular}{@{}lrr@{}}
\toprule
Variant & Best factor (DSR) & B\&H DSR \\
\midrule
Costs $\times 0.5$ & reversal (0.46) & 0.96 \\
Costs $\times 2$ & momentum (0.14) & 0.96 \\
Universe top-100 & momentum (0.62) & 0.73 \\
Universe top-300 & reversal (0.25) & 0.98 \\
\midrule
\multicolumn{3}{@{}l}{\emph{Multi-seed discovery (5 independent gpt-4.1 runs, $N{=}18$ each):}} \\
\multicolumn{3}{@{}l}{5/5 valid; all converge on the volatility-breakout family;} \\
\multicolumn{3}{@{}l}{eval SR $0.59$--$0.70$ ($+26$--$33\%$); DSR $0.15$--$0.29$; \textbf{0/5 survive}.} \\
\bottomrule
\end{tabular}
\end{table}

The null survives every assumption change we tested (Table~\ref{tab:e4}). Halving or doubling all cost components leaves every factor far from significance (best DSR $0.46$) while the passive benchmark's verdict is essentially unchanged (DSR $\approx 0.96$). Narrowing the point-in-time universe to the top 100 names makes momentum borderline-interesting (DSR $=0.62$, evaluation Sharpe $0.71$), still below the bar; widening to 300 collapses it (PBO $=0.91$). Most tellingly, five independent repetitions of the full autonomous protocol converge on the \emph{same} interpretable strategy family, all five profitable out of sample, and none of the five survives deflation. The discovery is reproducible; its evidence, at the agent's own search intensity over a four-year window, is insufficient, and reproducibly so. An independent re-evaluation of published LLM timing strategies over a longer horizon and a wider cross-section reports the same direction of result \citep{Li2025finsaber}, which is evidence that the pattern is not an artefact of this harness.

\subsection{E5: the fair field, and what the referee certifies}
\label{sec:e5}

If Sections~\ref{sec:e2}--\ref{sec:e4} were the whole story, the framework could be dismissed as an expensive proof of market efficiency. E5 moves to the arena where documented premia actually live, 39 multi-asset ETFs, and relaxes the constraints that suppress them: long/flat and long-only books are allowed, rebalancing is slow, and the evaluation window is nine years. A pre-registered human hypothesis (a divergence signal specified before any results were seen) enters as an arm alongside the agent's discovery, classic cross-asset momentum, time-series momentum \citep{Moskowitz2012tsm}, the random null, and passive benchmarks.

\begin{table}[pos=t]
\centering\small
\caption{E5 fair field (39 multi-asset ETFs; design 2007--2016, eval 2017--2025; ledger $N{=}20$, PBO $=0.80$). ``Certified'' = evaluation-window Sharpe CI excludes zero.}
\label{tab:e5}
\begin{tabular}{@{}lrrrl@{}}
\toprule
Arm & Des.\ SR & Eval SR & Eval ret. & Eval 95\% CI \\
\midrule
SPY buy-and-hold & 0.42 & 0.85 & $+249\%$ & $[+0.24, +1.53]$ \\
B\&H (39 ETFs) & 0.33 & 0.71 & $+108\%$ & $[+0.07, +1.46]$ \\
TSM long/flat & 0.42 & 0.49 & $+55\%$ & $[-0.13, +1.24]$ \\
XS momentum L/S & 0.24 & 0.20 & $+12\%$ & $[-0.39, +0.82]$ \\
Divergence (pre-reg.) & 0.34 & $-$0.15 & $-13\%$ & $[-0.73, +0.56]$ \\
Random floor & $-$0.57 & 0.55 & $+23\%$ & $[-0.06, +1.18]$ \\
Agent discovery & 0.37 & $-$0.23 & $-21\%$ & $[-0.82, +0.33]$ \\
\bottomrule
\end{tabular}
\end{table}

\begin{figure}[pos=t]
\centering
\includegraphics[width=\columnwidth]{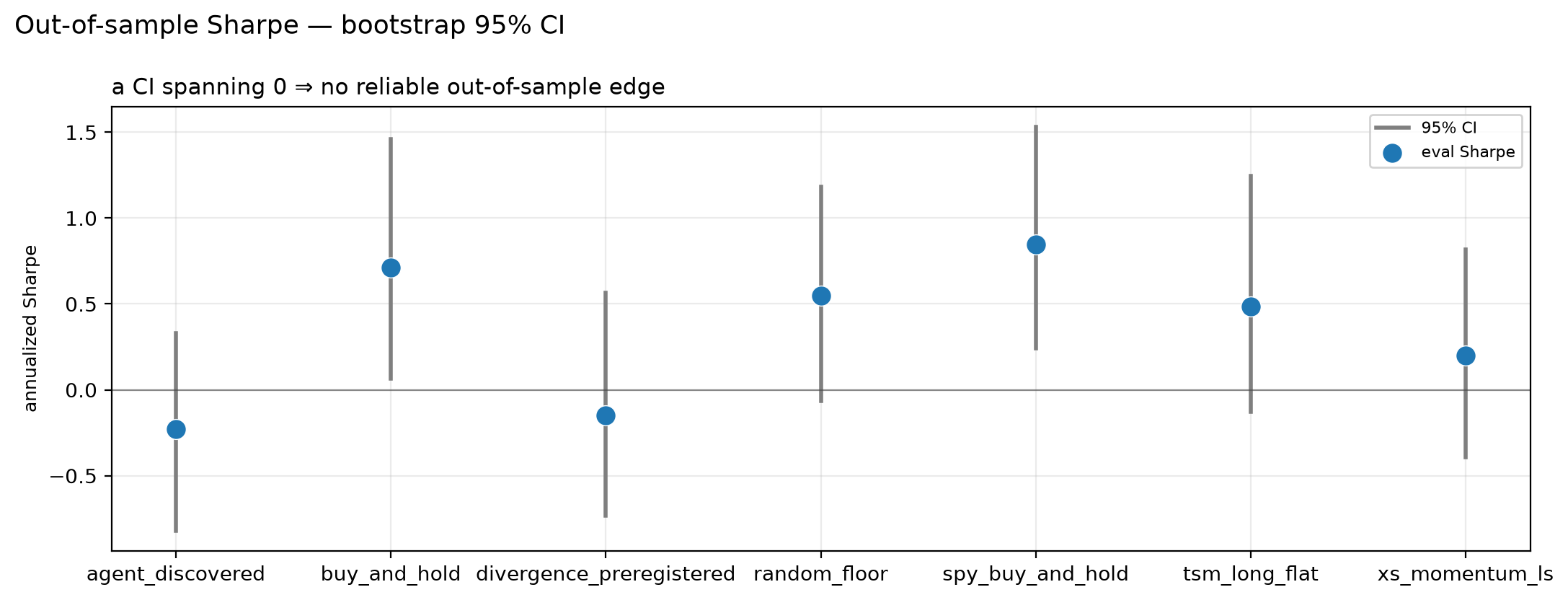}
\caption{E5: evaluation-window Sharpe with stationary-bootstrap 95\% intervals. Only the passive benchmarks exclude zero. A CI spanning zero is not evidence of absence; it is a statement about evidence, which is the point.}
\label{fig:ci-e5}
\end{figure}

Five findings (Table~\ref{tab:e5}, Fig.~\ref{fig:ci-e5}):

\begin{enumerate}
\item \textbf{The referee says yes when it should.} Equal-weight multi-asset buy-and-hold and SPY are certified. The instruments are not a universal rejector; they pass real premia and reject noise, both behaviors on the same table.
\item \textbf{The power arithmetic, made concrete.} Time-series momentum, among the most robust documented premia, is positive in both windows and returns $+55\%$ over nine held-out years, and is \emph{still} under-powered ($t \approx 0.49\sqrt{9} \approx 1.5$). Certifying moderate low-frequency edges requires decades or breadth; Section~\ref{sec:discussion} draws the consequences.
\item \textbf{Raw out-of-sample profit is not evidence.} The random-weights null made $+23\%$ over the nine evaluation years. Its verdict (design Sharpe $-0.57$, DSR $=0.0008$) is the cleanest illustration in the suite of why unaudited backtest profits, even ``out-of-sample'' ones, are not by themselves evidence of skill.
\item \textbf{PBO predicted the agent's failure in advance.} On this field the agent's best in-sample discovery ($+0.37$) degraded out of sample ($-0.23$); the search's PBO of $0.80$, computed on design data alone, said before the evaluation window was opened that the in-sample winner would likely underperform. Together with E3, the two distinct failure modes, uncertifiable selection luck and genuine overfitting, are each caught by the instrument built for them.
\item \textbf{Pre-registration earns a lower bar, not a free pass.} The human divergence hypothesis, entering at near-zero deflation, simply did not generalize cross-asset. The framework rewarded the epistemic posture and still reported the outcome.
\end{enumerate}

\subsection{E6/E6b: a human production system under the same instruments}
\label{sec:e6}

The pre-registered signal of E5/E6 was a simplification; the author's actual production system is a discrete weekly rule engine (PineScript) in which the divergence quantity serves as a \emph{hold/exit} test inside a trend-following state machine, with linear-regression slope gating, candle-anatomy filters, and all-in/all-out position state. E6b ports it faithfully: the system is expressed as an 84-node signal graph executed by the rule archetype (Section~\ref{sec:rulearch}), and the translation is cross-validated bar-for-bar against a reference implementation of the original script, with zero mismatching bars on all four entry/exit conditions, before any performance is examined. It enters the protocol as a pre-registered production system ($N{=}1$).

\begin{table}[pos=t]
\centering\small
\caption{E6b: the ported production system on gold (GLD; design 2007--2016, eval 2017--2025; $N{=}1$). $p$: paired bootstrap vs.\ holding gold.}
\label{tab:e6b}
\begin{tabular}{@{}lrrrr@{}}
\toprule
Arm & Des.\ SR / ret. & Eval SR / ret. & Eval CI & $p$ \\
\midrule
Gold buy-and-hold & 0.38 / $+75\%$ & 1.04 / $+257\%$ & $[+0.42,+1.62]$ & --- \\
SPY buy-and-hold & 0.42 / $+92\%$ & 0.85 / $+249\%$ & $[+0.24,+1.53]$ & 0.94 \\
Production system & 0.33 / $+34\%$ & $-0.12$ / $-13\%$ & $[-0.77,+0.49]$ & $<0.001$ \\
\bottomrule
\end{tabular}
\end{table}

The verdict is two-sided (Table~\ref{tab:e6b}). The system was genuinely profitable across its development decade: $+34\%$ through both a gold bull and the 2011--2015 bear, in a position only a third of all days, PSR $=0.85$ at the minimal deflation its pre-registration earned. It did not persist: over the nine evaluation years its flat and short states forfeited a historic bull market that buy-and-hold collected in full, and the paired test rejects it against holding gold. Notably, the author's own trading-platform backtest (zero costs, spot data, fixed-lot sizing) independently shows buy-and-hold finishing ahead of the strategy over the full sample, corroborating the benchmark-relative conclusion under a different implementation. Methodologically, E6b is the suite's completeness proof: a real human system, state machine and candle patterns included, runs under exactly the instruments that judge the LLM, with the same costs, the same windows, and the same verdicts.

\section{Discussion}
\label{sec:discussion}

\paragraph{The power arithmetic governs everything.}
A $t$-statistic on a Sharpe ratio grows as $SR\sqrt{\text{years}}$: a true $SR{=}0.6$ strategy needs roughly eleven years of out-of-sample data to reach $t \approx 2$, and deflation raises the bar further. This single fact organizes the results. It explains why nothing moderate certifies on four-year windows (E1--E4) and why even a robust premium at $+55\%$ over nine years does not (E5). It also explains, without mysticism, the canonical counterexamples to market efficiency: a high-frequency operation performing thousands of trades a day accumulates statistical power in months, and a long-only factor investor compounding for five decades accumulates it the slow way, consistent with the factor-based decompositions of such track records \citep{Frazzini2018buffett}. Non-certification is not non-existence; the referee reports evidence, not conviction, and quantifies exactly how much evidence a claimed edge still lacks.

\paragraph{Scaling the search.} Nothing in the architecture caps the budget at one hundred candidates, and larger budgets do not weaken the accounting; they tighten it. The ledger costs one row per trial, and the deflation threshold of Eq.~\ref{eq:dsr-threshold} grows with $\Phi^{-1}(1-1/N)$: at equal trial dispersion, a search of $10^{4}$ candidates faces a bar roughly $1.5\times$ higher than the hundred-candidate searches reported here. Under this accounting a bigger search cannot manufacture certification; it can only certify a discovery whose edge outruns the rising threshold, and the evaporation curve (Section~\ref{sec:evaporation}) makes exactly that race visible at any budget.

\paragraph{Three failure modes, three instruments.}
The suite's discoveries failed in three distinct ways. The multi-seed finds were profitable out of sample yet indistinguishable from selection luck at the agent's own trial count (a DSR failure); E5's find degraded out of sample exactly as its design-window PBO predicted (a CSCV success); and E3's large-budget find collapsed out of sample despite a \emph{low} PBO, because its winner kept its rank among mostly losing candidates while losing its absolute edge, which the held-out confidence interval caught. A single reported backtest number cannot express these distinctions, which is our argument for computing all three in an evaluation layer for strategy-discovering agents.

\paragraph{Structural guardrails over procedural promises.}
E1 quantifies why the two corrections cannot substitute for each other: a sufficiently contaminated information set survives statistical deflation entirely (DSR $=1.00$ at Sharpe 35). Symmetrically, no amount of leakage discipline corrects for search intensity. The engineering consequence is that both belong in the \emph{type system} of the platform, the agent-selectable feature registry and the single evaluation entry point, rather than in prompts, conventions, or reviewer trust. These comparisons double as component ablations: the JSON-critic baseline is the system without its validated action surface (0/60 usable candidates), the raw-versus-deflated columns of Tables~\ref{tab:e3}--\ref{tab:e5} are the system without search correction, and Eq.~\ref{eq:dsr-threshold} at $N{=}1$ is the system without the trial ledger. Each removal reverts to precisely the pathology it was built to prevent.

\paragraph{Beyond finance.}
Any agentic system that searches a solution space and self-reports its best result, such as code agents benchmarked on the tasks they iterated on or scientific-discovery agents reporting their best hypothesis, shares this structure: autonomy inflates the trial count, and honest reporting requires the search to be recorded and paid for. Trading is the domain where the accounting instruments already exist. We have not tested the design pattern demonstrated here (validated tool surface; complete trial ledger; deflation indexed to it) outside finance, but nothing in it is finance-specific. What replaces the Deflated Sharpe Ratio outside finance is the same correction in different clothes: a multiplicity adjustment indexed to the \emph{recorded} number of evaluations. For a code agent, that is performance on a sequestered test set judged against the number of candidate patches scored on the visible tests; for a discovery agent ranking hypotheses, false-discovery-rate control \citep{BenjaminiHochberg1995} over every hypothesis the search actually scored. DSR is finance's instantiation, an expected-maximum threshold specialized to non-normal return series, but the requirement it expresses is domain-independent: the trial count entering the correction must be the true one, which is precisely what a validated action surface feeding a complete ledger guarantees.

\section{Limitations}
\label{sec:limitations}

\emph{Survivorship.} The equity constituent list is a fixed current list; delisted names are absent except for eight that leave mid-sample. This flatters the active strategies (survivors trend upward), so it biases \emph{against} our null result rather than for it; the point-in-time liquidity selection removes the selection-timing component. A fully delisting-inclusive panel requires commercial data.
\emph{Feature scope.} News-sentiment and point-in-time fundamental features exist in the system but were excluded from the headline discovery runs for reproducibility (rate-limited and filer-dependent external sources); the price-feature action space bounds what the agent could find.
\emph{Model coverage.} Agent runs use \texttt{gpt-4.1} and \texttt{claude-sonnet-5}; the provider layer is model-agnostic, but we do not claim generality beyond these two models. Notably, default-temperature sampling converged within a model: three of five gpt-4.1 repetitions composed near-identical best strategies, so multi-seed dispersion understates search diversity.
\emph{Window dependence.} The 2017--2025 evaluation window contains a historic gold and equity bull; timing systems are structurally disadvantaged in such regimes (E5/E6b), which the paired tests report but cannot deconfound.
\emph{E6b fidelity.} The port uses rolling five-day weeks on exchange-hours ETF data with notional sizing; the original runs on calendar weeks, spot data, fixed lots, and zero costs. The translation is exact at the rule level (cross-validated), not at the market-microstructure level.
\emph{Statistical caveats.} DSR assumes trials are exchangeable draws; our hypothesis-rotation makes candidates heterogeneous, and duplicated candidates (the rotation wraps after twenty hints) are counted at face value; both choices are conservative for the agent. PBO with $N{\approx}20$ candidates is coarse.

\section{Conclusion}
\label{sec:conclusion}

We asked what remains of LLM-driven trading strategy discovery when the two corrections the empirical-finance literature has demanded for a decade, leakage-safety and search-aware deflation, are made structural properties of the discovery system itself. The answer is symmetric and, we argue, more useful than another reported Sharpe ratio. The instruments certify the arms with genuine out-of-sample evidence: passive risk premia pass with confidence intervals excluding zero. They withhold certification from what the evidence does not yet support: every LLM-discovered strategy in this suite, across two universes, two frontier models, cost regimes, search budgets up to one hundred candidates, and five repeated runs, including discoveries that were genuinely profitable out of sample, fails certification against the agent's own recorded trial count, with three distinct failure modes caught by the instruments built for them. They also expose a limit of statistical correction: a planted look-ahead oracle survives deflation at a Sharpe ratio of 35, which is our argument for treating leakage-safety as structural. And they extend the same hearing to a human production system, ported bar-for-bar and judged by identical rules.

The system functions as a certification mechanism: it builds strategies \emph{and} tells its operator which ones may be trusted, at what evidential price. The LLM's discoveries are real programs with real out-of-sample returns, and none of them, under honest accounting at the search intensity recorded here, is yet distinguishable from luck. That is not a failure of the system. It is the system working.

\section*{CRediT authorship contribution statement}
\textbf{Eray Gen\c{c}ay:} Conceptualization, Methodology, Software, Validation, Formal analysis, Investigation, Data curation, Writing -- original draft, Writing -- review \& editing, Visualization.

\section*{Funding}
This research did not receive any specific grant from funding agencies in the public, commercial, or not-for-profit sectors.

\section*{Declaration of competing interest}
The author declares that he has no known competing financial interests or personal relationships that could have appeared to influence the work reported in this paper.

\section*{Data and code availability}
A public archive containing the statistical evaluation code implementing the instruments of Section~\ref{sec:honest} (with its offline test suite), every experiment configuration, and the canonical run manifest and complete search trace behind each reported table is available on Zenodo.\footnote{DOI: \url{https://doi.org/10.5281/zenodo.21261868}} The full discovery platform is available from the author on reasonable request for research verification purposes. Raw market data were obtained from Tiingo, SEC EDGAR, GDELT, and FRED under their respective terms of use and cannot be redistributed by the author; the system includes resumable scripts that reconstruct the exact data panels from a free-tier Tiingo API key, and the manifests' content hashes allow independent verification that equivalent data were obtained.

\bibliographystyle{cas-model2-names}
\bibliography{references}

\end{document}